\documentclass[prd,twocolumn,showpacs,amsmath,amssymb,superscriptaddress,preprintnumbers,nofootinbib]{revtex4}
\usepackage{graphicx}
\usepackage{hyperref}

\normalbaselines
\newcommand{\D}{\hat{D}}
\newcommand{\cqg}{Classical Quantum Gravity\ }
\newcommand{\grg}{Gen. Relativ. Gravit.\ }

\begin{document}

\title{Dark energy fingerprints in the nonminimal Wu-Yang wormhole structure}
\author{Alexander B. Balakin}
\email{Alexander.Balakin@kpfu.ru} \affiliation{Department of
General Relativity and Gravitation, Institute of Physics, Kazan
Federal University, Kremlevskaya str. 18, Kazan 420008, Russia}

\author{Alexei E. Zayats}
\email{Alexei.Zayats@kpfu.ru} \affiliation{Department of General
Relativity and Gravitation, Institute of Physics, Kazan Federal
University, Kremlevskaya str. 18, Kazan 420008, Russia}
%
%\preprint{\bf version 2.2}
%
%
%\date{\today}%

\begin{abstract}
We discuss new exact solutions to nonminimally extended
Einstein-Yang-Mills equations describing  spherically symmetric
static wormholes supported by the gauge field of the Wu-Yang type
in a dark energy environment. We focus on the analysis of three
types of  exact solutions to the gravitational field equations.
Solutions of the first type relate to the model, in which the dark
energy is anisotropic, i.e., the radial and tangential pressures
do not coincide. Solutions of the second type correspond to the
isotropic pressure tensor; in particular, we discuss the exact
solution, for which the dark energy is characterized by the
equation of state for a string gas. Solutions of the third type
describe the dark energy model with constant pressure and energy
density. For the solutions of the third type, we consider in
detail the problem of horizons and find constraints for the
parameters of nonminimal coupling and for the constitutive
parameters of the dark energy equation of state, which guarantee
that the nonminimal wormholes are traversable.
\end{abstract}


\maketitle

\section{Introduction}

Dark energy, the main constituent of a cosmic dark fluid, is
considered nowadays as a key element of numerous cosmological
models (see, e.g., \cite{DE1,DE2,DE3,DE5}). Originally, the term
dark energy was introduced into the scientific lexicon in order to
explain the discovery of the late-time accelerated expansion of
the Universe \cite{AE1,AE2,AE3}. However, there exist alternative
explanations of this observational fact, for instance, in the
framework of modified gravity (see, e.g., the review \cite{DE4}
for references and details). This means that we need to extend our
knowledge about interactions of the dark energy with matter and
fields in order to say definitely whether the dark energy is the
real medium with exotic properties or the accelerated expansion of
the Universe is the effective result of gravitational interactions
with modified laws of attraction/repulsion. Clearly, there are
dark energy interactions of two types: the indirect and direct
ones. The first channel of indirect interaction with matter and
fields is realized via the global gravity field; this channel is
just visualized by the accelerated expansion of the Universe. The
most known channel of direct interaction is presented by the
coupling of the dark energy with dark matter, the second
constituent of the dark fluid (see, e.g.,
\cite{DEDM1,DEDM2,DEDM3}). The interactions of this type describe
fine details of the cosmological expansion thus clarifying the
fate of the Universe (big rip, little rip, pseudo-rip, etc.),
solving the coincidence problem and answering the question of how
many epochs of accelerated and decelerated expansion the Universe
history includes. There are a few mathematical schemes of the
description of the coupling between the dark energy and dark
matter; for instance, in \cite{DEDM7} we introduced the model
force linear in the four-gradient of the dark energy pressure,
which acts on the dark matter particles in analogy with the
classical Archimedean force. The theory of direct dark energy
interactions with scalar, pseudoscalar, vector, electromagnetic,
and gauge fields is less elaborated at present but seems to be
very promising from the point of view of cosmological and
astrophysical applications. For instance, in \cite{BD2014}
studying the pyromagnetic, piezoelectric, and striction-type
schemes of the direct interactions between the dark energy and
electromagnetic field, we have found that specific unlighted
epochs can appear in the Universe history. In other words, one can
try to find some dark energy fingerprints in the Universe history,
which are marked due to the coupling with electromagnetic fields.

In this paper we consider the nonminimal scheme of the indirect
coupling of the dark energy to the gauge field. The model with an
${\rm SU(N)}$ symmetric gauge field is known to be indicated as
the nonminimal one, if the Lagrangian of the model contains the
so-called cross-invariants constructed as a tensorial product of
the Riemann tensor, $R^i_{\ kmn}$, and its convolutions, on the
one hand, and of the Yang-Mills field strength tensor,
$F^{(a)}_{ik}$, on the other hand (see, e.g., \cite{NM1,NM2,NM3}
for references). In order to interpret the role of the dark energy
in such models, let us focus on the examples of static spherically
symmetric solutions to the nonminimally extended
Einstein-Yang-Mills equations (see \cite{NM3,EYM1,EYM2} for
details). When the dark energy is considered to be absent, we
refer to the exact solutions describing regular nonminimal Wu-Yang
monopoles and wormholes of a magnetic type \cite{EYM1,EYM2} as
well as traversable nonminimal wormholes of an electric type
\cite{BLZ10}. When the dark energy appears as the third player in
the nonminimal model, we intend to focus on the search for dark
energy fingerprints in the causal structure of the mentioned
spherically symmetric static objects. To be more precise, in this
paper, we focus on exact solutions of the nonminimally extended
Einstein-Yang-Mills model, which describe nonminimal wormholes
supported by Yang-Mills field in the dark energy environment.

Let us mention that in the cosmological context the dark energy is
usually considered as a spatially homogeneous substrate, the
pressure and energy density of which depend on time only. The most
known model of this class relates to the so-called
$\Lambda$-representation of the dark energy, characterized by
constant energy density $W_{({\rm DE})} = \frac{\Lambda}{8\pi}$
and pressure $P_{({\rm DE})} = -\frac{\Lambda}{8\pi}$ (several
examples of wormhole solutions with the $\Lambda$-term can be
found, e.g., in \cite{Kim,Li,LLO}). However, when we deal with the
dark energy influence on the throat structure of the spherically
symmetric static nonminimal wormhole, it is natural to treat the
dark energy as a static spatially inhomogeneous substrate, the
pressure of which depends on the radial variable $r$ only. The
corresponding approach is motivated mathematically in
Subsection~\ref{DEassump}.

The paper is organized as follows. In Section~\ref{keyeqs}, we
formulate the six-parameter nonminimal Einstein-Yang-Mills-dark
energy model, and represent master equations for the gauge and
gravitational fields. In Section~\ref{WYs}, we introduce key
assumptions about the equation of state of the dark energy and
describe the Wu-Yang ansatz for the structure of the gauge field.
In Section~\ref{Sect}, we present exact solutions of the wormhole
type for various models of the dark energy, including the
$\Lambda$-term-type configuration. In Section~\ref{Causal}, we
discuss in detail the causal structure of the spacetime for the
$\Lambda$-type case. Section~\ref{conclusion} is devoted to
discussions.

\section{Nonminimal master equations}\label{keyeqs}

\subsection{The action functional}

We start with the action functional
\begin{align}
S =& \int d^4 x \sqrt{{-}g} \left[\frac{R}{16\pi}+ L_{({\rm
DE})}+ \frac{1}{4}F^{(a)}_{ik} F^{ik(a)}\right.\nonumber\\
+&\left.{}\frac{1}{4} {\cal R}^{ikmn}F^{(a)}_{ik} F^{(a)}_{mn}
\right],\label{act}
\end{align}
where as usual $g = {\rm det}(g_{ik})$ is the determinant of a
metric tensor $g_{ik}$, $R$ is the Ricci scalar, the Latin indices
without parentheses run from 0 to 3. The term $L_{({\rm DE})}$ is
the Lagrangian describing the dark energy. The tensor of
nonminimal susceptibility ${\cal R}^{ikmn}$ is defined as follows:
\begin{align}
{\cal R}^{ikmn} &\equiv
\frac{q_1}{2}R\,(g^{im}g^{kn}-g^{in}g^{km}) \nonumber\\
&{}+ \frac{q_2}{2}(R^{im}g^{kn} -R^{in}g^{km}+ R^{kn}g^{im} -R^{km}g^{in}) \nonumber\\
&{}+ q_3 R^{ikmn}\,, \label{sus}
\end{align}
where $R^{ik}$ and $R^{ikmn}$ are the Ricci and Riemann tensors,
respectively, and $q_1$, $q_2$, $q_3$ are the phenomeno\-logi\-cal
parameters describing the nonminimal coupling of the Yang-Mills
fields with gravitation. We consider the Yang-Mills fields taking
the values in the Lie algebra of the gauge group ${\rm SU}(2)$, so
that $A^{(a)}_i$ and $F^{(a)}_{mn}$ are the Yang-Mills field
potential and strength, respectively, the group index $(a)$ runs
from 1 to 3. The Yang-Mills fields $F^{(a)}_{mn}$ are connected
with the potentials of the gauge field $A^{(a)}_i$ by the
well-known formulas (see, e.g., \cite{Rubakov})
\begin{equation}
F^{(a)}_{mn} = \nabla_m A^{(a)}_n - \nabla_n A^{(a)}_m +
f^{(a)}_{\ \ (b)(c)} A^{(b)}_m A^{(c)}_n \,. \label{Fmn}
\end{equation}
Here $\nabla _m$ is a  covariant spacetime derivative, the
sym\-bols $f_{(a)(b)(c)}\equiv \varepsilon_{(a)(b)(c)}$ denote the
real structure constants of the gauge group ${\rm SU}(2)$.

The nonminimal susceptibility tensors ${\cal R}^{ikmn}$
(\ref{sus}) contains three phenomenological parameters $q_1$,
$q_2$ and $q_3$. We consider these three parameters to be
independent coupling constants, and this choice is historically
motivated. In the pioneer work \cite{Prasanna71}, Prasanna
introduced only one phenomenological parameter in front of the
Riemann tensor $R^{ikmn}$ in the new cross-term $q
R^{ikmn}F_{ik}F_{mn}$ appeared in the nonminimally extended
Lagrangian. Later, Drummond and Hathrell \cite{DH}, using the
one-loop corrections to QED, showed that the tensor ${\cal
R}^{ikmn}$ possesses just the structure (\ref{sus}),  with $q_1
=-5q$, $q_2=13q$, $q_3=-2q$, where the positive parameter $q
\equiv \frac{\alpha\lambda^2_{\rm e}}{180\pi}$ is constructed by
using the fine structure constant $\alpha$ and the Compton
wavelength of the electron $\lambda_{\rm e}$. In other words,
direct calculations of Drummond and Hathrell have fixed attention
on the fact that there are three different coupling nonminimal
constants, $q_1$, $q_2$, and $q_3$, which are proportional to the
one nonminimal parameter $q$ with the dimensionality of length in
square. In the paper \cite{BL05}, a general Einstein-Maxwell model
was studied, in the framework of which the coupling constants
$q_1$, $q_2$ and $q_3$ are considered to be independent
parameters. The motivation of this idea is based on the
irreducible representation of the curvature tensor
\begin{equation}\label{irredu3}
 R^{ikmn} = {\cal C}^{ikmn} + {\cal E}^{ikmn} +  {\cal G}^{ikmn}  \,,
\end{equation}
where ${\cal C}^{ikmn}$ is the traceless Weyl tensor, and the
standard formulas are introduced:
\begin{gather}
{\cal E}^{ikmn} \equiv \frac{1}{2} \left({\cal S}^{im}g^{kn} {-}
{\cal S}^{in}g^{km} {+} {\cal S}^{kn}g^{im} {-} {\cal
S}^{km}g^{in}) \right) \,, \nonumber\\
{\cal S}^{mn} \equiv R^{mn}{-}
\frac{1}{4} R g^{mn} \,,\nonumber\\
{\cal G}^{ikmn} \equiv \frac{1}{12} R \ (g^{im}g^{kn} {-}
g^{in}g^{km}) \,, \nonumber\\ {\cal C}^{m}_{\ \ nmk} =0 \,, \quad
{\cal S}^{m}_m=0 \,, \quad {\cal G}^{mn}_{\ \ mn} =
R\label{irredu2}
\end{gather}
(we follow the notations from the book \cite{ExactSol}). Since
${\cal C}^{ikmn}$, ${\cal E}^{ikmn}$, and ${\cal G}^{ikmn}$ are
independent (irreducible) parts of the decomposition of the
Riemann tensor, it is reasonable to represent the nonminimal
susceptibility tensor as the sum of three independent parts:
\begin{equation}\label{irredu}
{\cal R}^{ikmn} {=} \lambda_1{\cal G}^{ikmn} {+} \lambda_2 {\cal
E}^{ikmn} {+}  \lambda_3 {\cal C}^{ikmn}  \,,
\end{equation}
with phenomenological constants $\lambda_1$, $\lambda_2$ and
$\lambda_3$. Clearly, (\ref{irredu}) converts into (\ref{sus}),
when
\begin{equation}\label{irredu7}
\lambda_1 = 6q_1 {+} 3q_2 {+} q_3 \,, \quad \lambda_2 = q_2 {+}
q_3 \,, \quad \lambda_3 = q_3\,.
\end{equation}
In other words, we have two equivalent irreducible decompositions
of the susceptibility tensor, (\ref{irredu}) and (\ref{sus}), but
we prefer to use (\ref{sus}), keeping in mind historical motives.

\subsection{Nonminimal extension of the Yang-Mills equations}
Variation of the action (\ref{act}) with respect to the Yang-Mills
potential $A_i^{(a)}$ yields
\begin{equation}
\nabla_k H^{(a)ik}+ f^{(a)}_{\ \ (b)(c)}
A^{(b)}_k  H^{(c)ik} = 0 \,. \label{YMeq}
\end{equation}
The tensor $H^{(b)ik} = F^{(b)ik} + {\cal R}^{ikmn} F^{(b)}_{mn}$ is a non-Abelian analog of the excitation tensor, known in
the electrodynamics \cite{Maugin,HO}.

\subsection{Master equations for the gravitational field}
The variation of the action functional (\ref{act}) with respect to
metric yields
\begin{equation}\label{EinMaster}
R_{ik}-\frac{1}{2}Rg_{ik}= 8\pi T^{\rm (eff)}_{ik}\,.
\end{equation}
The effective stress-energy tensor $T^{({\rm eff})}_{ik}$  can be
divided into five parts:
\begin{gather}
T^{\rm (eff)}_{ik} = T^{({\rm DE})}_{ik} +  T^{({\rm YM})}_{ik}
\nonumber\\ {}+ q_1 T^{(I)}_{ik} + q_2 T^{(II)}_{ik} + q_3
T^{(III)}_{ik} \,. \label{Tdecomp}
\end{gather}
The first term
\begin{equation}
T^{({\rm DE})}_{ik} \equiv -\frac{2}{\sqrt{-g}} \frac{\delta
\left[\sqrt{-g} L_{({\rm DE})}\right]}{\delta g^{ik}} \,,
\label{TDE}
\end{equation}
describes the stress-energy tensor of the dark energy. As usual,
we assume that this tensor possesses timelike eigen-four-vector
$U^i$ normalized by unity and denote the corresponding eigenvalue
as $W$, i.e., we assume that
\begin{equation}
T^{({\rm DE})}_{ik} U^k = W U_i \,, \quad g_{ik}U^iU^k = 1 \,. \label{eigen}
\end{equation}
The other three eigenvalues are denoted as $\Pi_1$, $\Pi_2$, and
$\Pi_3$. The second term
\begin{equation}
T^{({\rm YM})}_{ik} \equiv \frac{1}{4} g_{ik} F^{(a)}_{mn}F^{mn(a)} -
F^{(a)}_{in}F_{k}^{\ n(a)} \,, \label{TYM}
\end{equation}
is a stress-energy tensor of the pure Yang-Mills field. The last
three terms in (\ref{Tdecomp}) describe nonminimal contributions
into the stress-energy tensor. Since the parameters $q_1$, $q_2$,
and $q_3$ are independent coupling constants appearing in the
irreducible representation of the susceptibility tensor ${\cal
R}^{ikmn}$ (\ref{sus}), we decomposed this nonminimal contribution
into the stress-energy tensor as a sum of three terms with $q_1$,
$q_2$, and $q_3$ in front of the following tensors:
\begin{align}%
T^{(I)}_{ik} &= R\,T^{(YM)}_{ik} -  \frac{1}{2} R_{ik}
F^{(a)}_{mn}F^{mn(a)} \nonumber\\
&{}+\frac{1}{2} \left[ {\D}_{i} {\D}_{k} - g_{ik} {\D}^l {\D}_l
\right] \left[F^{(a)}_{mn}F^{mn(a)} \right], \label{TI}
\end{align}%
\begin{align}%
T^{(II)}_{ik} &= -\frac{1}{2}g_{ik}\biggl[{\D}_{m}
{\D}_{l}\left(F^{mn(a)}F^{l\ (a)}_{\ n}\right)
\nonumber\\
&{}-R_{lm}F^{mn (a)} F^{l\ (a)}_{\ n} \biggr] {-} F^{ln(a)}
\left(R_{il}F^{(a)}_{kn} + R_{kl}F^{(a)}_{in}\right)\nonumber\\
&{}- R^{mn}F^{(a)}_{im} F_{kn}^{(a)} - \frac{1}{2} {\D}^m{\D}_m
\left(F^{(a)}_{in} F_{k}^{ \
n(a)}\right)\nonumber\\
&{}+\frac{1}{2}{\D}_l \left[ {\D}_i \left( F^{(a)}_{kn}F^{ln(a)}
\right) + {\D}_k \left(F^{(a)}_{in}F^{ln(a)} \right) \right],
\label{TII}
\end{align}%
\begin{align}%
T^{(III)}_{ik} &= \frac{1}{4}g_{ik}
R^{mnls}F^{(a)}_{mn}F_{ls}^{(a)} \nonumber\\ &{}- \frac{3}{4}
F^{ls(a)} \left(F_{i}^{\ n(a)} R_{knls} + F_{k}^{\
n(a)}R_{inls}\right) \nonumber\\ &{}-\frac{1}{2}{\D}_{m} {\D}_{n}
\left[ F_{i}^{ \ n (a)}F_{k}^{ \ m(a)} + F_{k}^{ \ n(a)} F_{i}^{ \
m(a)} \right]. \label{TIII}
\end{align}%
We use the following rule:
\begin{align}
\D_m Q^{(a) \cdot \cdot \cdot}_{\cdot \cdot \cdot (d)} &\equiv
\nabla_m Q^{(a) \cdot \cdot \cdot}_{\cdot \cdot \cdot (d)} +
f^{(a)}_{\cdot (b)(c)} A^{(b)}_m Q^{(c) \cdot \cdot \cdot}_{\cdot
\cdot \cdot (d)}+\ldots \nonumber\\ &{}-  f^{(c)}_{\cdot (b)(d)}
A^{(b)}_m Q^{(a) \cdot \cdot \cdot}_{\cdot \cdot \cdot (c)}
-\ldots  \label{DQ2}
\end{align}
for the derivative of the arbitrary tensor defined in the group
space \cite{Akhiezer}.

\subsection{Compatibility conditions}

The Bianchi identities require the total stress-energy tensor to
be divergence-free, i.e., $\nabla^k T^{\rm (eff)}_{ik}{=}0$. Using
the decomposition (\ref{Tdecomp})-(\ref{DQ2}) and the Yang-Mills
field equations (\ref{YMeq}) one can show explicitly that this
compatibility condition reduces to the requirement $\nabla^k
T^{({\rm DE})}_{ik}=0$, i.e., we deal with a separate conservation
law for the dark energy. When all four eigenvalues, $W$, $\Pi_1$,
$\Pi_2$, and $\Pi_3$ coincide, one obtains that $T^{({\rm
DE})}_{ik}= \frac{\Lambda}{8\pi}\, g_{ik}$, and the compatibility
conditions yield $\nabla_i \Lambda =0$, i.e., $\Lambda$ is a
constant. In other words, the case when the dark energy can be
presented in terms of cosmological constant,
\begin{equation}\label{lambda}
W=\frac{\Lambda}{8\pi} \,, \quad \Pi_1 = \Pi_2 = \Pi_3 = - P =
\frac{\Lambda}{8\pi} = W \,,
\end{equation}
is also included into our scheme of analysis.

\section{Nonminimal wormhole of the Wu-Yang type in the dark energy environment}\label{WYs}

In order to model a traversable wormhole as a bridge joining two
different spacetimes, one needs to have specific ``bricks''
possessing exotic properties, which guarantee violation of the
null energy condition in the wormhole throat \cite{MorTho,HocVis}.
The history of investigations in the wormhole sector of field
theory can be found in \cite{Visserbook,Lobo}; we attract the
attention of the reader to the two episodes only. In 1973, Ellis
\cite{Ell} and, independently, Bronnikov \cite{Bro:73} constructed
wormhole solutions in the framework of the Einstein theory of
gravity with a scalar field, which has negatively defined kinetic
energy ({\it phantom} field in modern terminology). This first
example can be indicated, indeed, as the exotic one from the point
of view of classical field theory. The second example is connected
with the work of Barcell\'o and Visser \cite{BarVis:00}, who have
shown that an ordinary (standard) scalar field can support a
wormhole structure if this scalar field is nonminimally coupled to
the spacetime curvature. These two examples illustrate the typical
alternative: in order to justify the wormhole existence one can
either provide exotic properties of the ``bricks'' for its throat
or admit that the ``bricks'' are ordinary, but they interact
nonminimally with spacetime curvature. We use the second way for
modeling the wormhole configurations. For instance, in
\cite{BLZ10} one can find exact solutions for a traversable
wormhole supported by an electric field nonminimally coupled to
curvature. In the paper \cite{EYM2}, we reconstructed the wormhole
of the Wu-Yang type supported by SU(2) symmetric gauge field of
the magnetic type nonminimally coupled to gravity. Since the model
that we study now is a natural extension of that model, let us
comment briefly three key-elements of the wormhole-type solutions
with a Wu-Yang-type  gauge field nonminimally coupled to gravity
in case when the interactions with dark energy are absent.

First of all, from our point of view the gauge fields attract
special attention in the context of wormhole modeling: on the one
hand, the Yang-Mills field is the most known contributor into the
modern High-Energy Physics models; on the other hand, the
Yang-Mills potentials form SU(N) multiplets, thus providing a
versatile instrument for theoretical modeling.

Second, the pure Yang-Mills field itself is nonexotic, since it
cannot violate the null energy condition in the wormhole throat,
and, therefore, known wormhole solutions require either an
additional {\it phantom} scalar field \cite{Hauser} or a surgery
technique \cite{Mazh}; however, being coupled nonminimally to the
spacetime curvature, the Yang-Mills field becomes able to organize
a traversable wormhole throat (see \cite{EYM2}).

Third, the traversability of the Wu-Yang wormhole and the value of
the radius of the throat can be the subject of tuning: depending
on the relationships between the principal nonminimal coupling
constant $q$ and the magnetic charge $\nu$, the throat happens to
be closed or opened (see \cite{EYM2} for details). The following
question is of great interest:  physical processes of what type
could regulate the traversability of the nonminimal Wu-Yang
wormhole? The results of the paper \cite{EYM2} show that when the
charge $\nu$ grows horizons appear and the wormhole becomes
nontraversable (the cosmic gates happen to be closed). When the
coupling constant $q$ increases, the cosmic gates, per contra,
become opened. As it will be shown below, when we consider an
additional element of the theoretical modeling, the dark energy,
we obtain a new possibility for modeling the cosmic gates opening
and closing.

\subsection{Wu-Yang type solution to the Yang-Mills field equations}

Let us consider a static spherically symmetric spacetime with the
metric
\begin{gather}
ds^2=\sigma^2Ndt^2-\frac{dr^2}{N}- R^2(r) \left( d\theta^2 +
\sin^2\theta d\varphi^2 \right) ,\nonumber\\
r\in(-\infty;+\infty)\label{metrica}
\end{gather}
Here $\sigma$, $N$, and $R(r)$ are functions depending on the
radial coordinate $r$ only. Since the functions, which
characterize the dark energy, do not enter the master equations
for the gauge field, we obtain immediately that, as in the cases
of a nonminimal ${\rm SU}(2)$ monopole \cite{EYM1} and wormhole
\cite{EYM2}, the special ansatz (see \cite{Bais,WuYang}),
\begin{equation}\label{1}
\mathbf{A}_{0}=\mathbf{A}_{r}=0 \,,\quad
\mathbf{A}_{\theta}=i\mathbf{t}_{\varphi},\quad
\mathbf{A}_{\varphi}=-i\nu\sin{\theta}\;\mathbf{t}_{\theta}\,,
\end{equation}
gives the exact solution with the field strength tensor
of the following form:
\begin{equation}\label{2}
{\bf F}_{ik} = \delta_i^{\theta} \delta_k^{\varphi}{\bf F}_{\theta\varphi} \,, \quad {\bf F}_{\theta\varphi}=i\nu\sin\theta\,{\bf t}_{r}\,.
\end{equation}
The parameter $\nu$ is an integer and it denotes a magnetic
charge. Let us remind that the generators ${\bf t}_r$, ${\bf
t}_{\theta}$ and ${\bf t}_{\varphi}$ are the position-dependent
ones and are connected with the standard generators of the ${\rm
SU}(2)$ group as
\begin{gather}%
{\bf t}_r=\cos{\nu\varphi} \ \sin{\theta}\;{\bf
t}_{(1)}+\sin{\nu\varphi} \ \sin{\theta}\;{\bf
t}_{(2)}+\cos{\theta}\;{\bf t}_{(3)},\nonumber \\ {\bf
t}_{\theta}=\partial_{\theta}{\bf t}_r,\qquad {\bf
t}_{\varphi}=\frac {1}{\nu\sin{\theta}}\ \partial_{\varphi}{\bf
t}_r ; \label{deS5}
\end{gather}%
they satisfy the relations
\begin{equation}%
\left[{\bf t}_{r},{\bf t}_{\theta}\right]=i\,{\bf t}_{\varphi}
\,,\quad \left[{\bf t}_{\theta} \,, {\bf
t}_{\varphi}\right]=i\,{\bf t}_{r} \,, \quad \left[{\bf
t}_{\varphi},{\bf t}_{r}\right]=i\,{\bf t}_{\theta}\,.\label{deS6}
\end{equation}%
The system of Yang-Mills equations (\ref{YMeq}) is satisfied identically for arbitrary curvature tensor
and for arbitrary equation of state of the dark energy.

Again, this solution is effectively Abelian, i.e., by the suitable
gauge transformation ${\bf U}=\exp(-i\,{\bf \theta \, t_\varphi})$
it can be converted into the product of the Dirac-type potential
and the gauge group generator ${\bf t}_{(3)}$.

\subsection{Assumptions about the dark energy}\label{DEassump}

The simplest variant to introduce the dark energy is known to
connect with cosmological $\Lambda$ term. In the de~Sitter
spacetime with the so-called $t$-representation of the metric
\begin{equation}\label{ds1}
 ds^2= dt^2 - a^2(t_0)e^{2\sqrt{\frac{\Lambda}{3}}(t-t_0)}\left(dr^2 + r^2 d\Omega^2  \right)  \,,
\end{equation}
the dark energy pressure and energy-density scalar are constant
and they can be written, respectively, in the form
\begin{gather}
P_{({\rm DE})} = - \frac{\Lambda}{8\pi} \,, \quad  W_{({\rm DE})}
=  \frac{\Lambda}{8\pi} \,,\nonumber \\ \quad P_{({\rm DE})} +
W_{({\rm DE})} = 0 \,.\label{ds2}
\end{gather}
In more sophisticated models these quantities are considered to be
functions of cosmological time  $P_{({\rm DE})}(t)$, $W_{({\rm
DE})}(t)$, nevertheless, the specific equation of state with
$P_{({\rm DE})}(t) + W_{({\rm DE})}(t) \leq 0$ remains the
distinguishing feature.

The de~Sitter spacetime can also be described in the so-called
$r$-representation using the equivalent metric
\begin{equation}\label{ds3}
 ds^2= \left(1- \frac{\Lambda r^2}{3} \right) dt^2 - \left(1- \frac{\Lambda r^2}{3} \right)^{-1} dr^2 - r^2 d\Omega^2  \,,
\end{equation}
which, clearly, is static and depends on the radial variable $r$
only. The corresponding Schwarzschild-Reissner-Nordstr\"om, etc.,
extensions of the model lead to the replacements
\begin{gather}
 \left(1- \frac{\Lambda r^2}{3} \right) \ \to \ \left(1- \frac{\Lambda r^2}{3} - \frac{2M}{r}  + \frac{Q^2}{2r^2} \right)
 \ \to \
 \nonumber\\ \to \ \left(1- \frac{\Lambda r^2}{3} + f(r) \right)  \,,
\label{ds4}
\end{gather}
where $M$ is the mass and $Q$ is the charge of the object.
Studying this model and more sophisticated models, we have a
possibility to interpret the contributions containing $\Lambda$ in
terms of dark energy of the $\Lambda$-type. Our ansatz is that one
can develop this idea and consider the dark energy in the
$r$-representation, i.e., using the functions $P_{({\rm DE})}(r)$,
$W_{({\rm DE})}(r)$ satisfying the typical condition $P_{({\rm
DE})}(r) + W_{({\rm DE})}(r) \leq 0$.

For the static spherically symmetric configuration, the two
diagonal components of the effective stress-energy tensor
coincide, ${T^{({\rm eff})}}^{\theta}_{\theta} {=}{T^{({\rm
eff})}}^{\varphi}_{\varphi}$. Based on this fact, we assume that
two of the three eigenvalues $\Pi_1$, $\Pi_2$, $\Pi_3$ of the
stress-energy tensor of the dark energy also coincide, thus
providing the following definitions
\begin{equation}
- P_{||} \equiv {T^{({\rm DE})}}^{r}_{r} \,, \quad - P_{\bot} \equiv  {T^{({\rm DE})}}^{\theta}_{\theta} {=}{T^{({\rm DE})}}^{\varphi}_{\varphi}
 \,.
\label{eos1}
\end{equation}
In other words we can introduce longitudinal, $P_{||}$, and transversal, $P_{\bot}$ pressures, prescribed to the dark energy.
We assume that the dark energy equations of state are linear
\begin{equation}
P_{||} = \omega_{||} \ W \,, \quad P_{\bot} = \omega_{\bot} W \,.
\label{eos3}
\end{equation}
When the dark energy is considered to possess an isotropic
pressure, we have to put $\omega_{||} = \omega_{\bot}$.

\section{Exact solutions to the gravitational field equations}\label{Sect}

\subsection{Key equations}

For the metric (\ref{metrica}), only four components of the
Einstein tensor $ G_i^{\,k}{=}R_i^{\,k}{-}\frac{1}{2}\delta_i^k R$
are nonvanishing, $G^0_0$, $G^r_r$, and $G^{\theta}_{\theta} {=}
G^{\varphi}_{\varphi}$. In order to describe the gravity field we
use the following three independent equations. First, we consider
the difference of the first two equations, $G^0_0{-}G^r_r {=} 8\pi
({T^{({\rm eff})}}^0_0{-}{T^{({\rm eff})}}^r_r)$, yielding
\begin{gather}
\left(1-\frac{\kappa q_1}{R^4}\right)\left[\frac{\sigma'R'}{\sigma
R}-\frac{R''}{R}\right]\nonumber\\
{}=\frac{\kappa(10q_1+4q_2+q_3)R'^2}{R^6}+\frac{4\pi(W+P_{||})}{N}.
\label{eq1}
\end{gather}
Second, we use the Einstein equation $G^0_0{=}8\pi {T^{({\rm
eff})}}^0_0$, which gives
\begin{gather}
\frac{1-NR'^2}{R^2}-\left(1-\frac{\kappa
q_1}{R^4}\right)\left(\frac{N'R'}{R}+\frac{2NR''}{R}\right)\nonumber\\
{}=\frac{\kappa}{R^4}\left\{\frac12{-}\frac{q_1{+}q_2{+}q_3}{R^2}{+}\frac{(13q_1{+}4q_2{+}q_3)NR'^2}{R^2}\right\}\nonumber\\
{}+8\pi W \,. \label{eq2}
\end{gather}
Third, we consider the compatibility equation $\nabla^k T^{({\rm
eff})}_{ik}{=}0$, which can be reduced now to the equation of
hydrostatic equilibrium
\begin{equation}
P^{\prime}_{||} + \frac{2R^{\prime}}{R} \left(P_{||}-P_{\bot} \right) +
\left(W+ P_{||} \right) \frac{\left(\sigma^2 N \right)^{\prime}}{2\sigma^2 N} = 0
 \,.
\label{eos2}
\end{equation}
The parameter $\kappa$ is defined as $\kappa=8\pi\nu^2$. The prime
denotes the derivative with respect to the variable $r$.

\subsubsection{Energy density distribution}

The equation (\ref{eos2}) can be easily resolved using equations
of state (\ref{eos3}); this procedure yields
\begin{gather}
W(r) = W_0  \left[\frac{R(r)}{R(0)}\right]^{2 \alpha}
\left[\sigma^2(r) N(r) \right]^{- \gamma}, \nonumber\\ \alpha
\equiv \frac{\omega_{\bot}-\omega_{||}}{\omega_{||}} , \quad
\gamma \equiv \frac{1+\omega_{||}}{2 \omega_{||}}, \label{eos4}
\end{gather}
where $R(0) \neq 0$ is the value of the radial function $R(r)$ at
$r=0$ and $W_0$ is an integration constant with dimensionality of
energy density. We have to stress, that the formula (\ref{eos4})
describes the distribution of the energy density of the dark
energy for arbitrary radial function $R(r)$ satisfying the
condition $R(0) \neq 0$.

\subsubsection{Examples of the radial function $R(r)$}

The function $R(r)$ can be chosen according to physical
requirements; when we consider the wormhole-type solutions, we
assume that
\begin{equation}
R(0)=a>0 \,, \quad R'(0)=0\,, \quad R''(0)>0 \,, \label{rrr}
\end{equation}
where $a$ is the throat radius:

\noindent {\it (1)} \noindent The most known function satisfying
these conditions is $R(r) {=} \sqrt{r^2 {+}a^2}$; it is even with
respect to the variable $r$. We will use it below for the
reconstruction of exact solutions to the master equations with
constant dark energy pressure; this radial function displays the
asymptotic behavior $R(r\to \infty) \to r$.

\noindent
{\it (2)}
\noindent The conditions (\ref{rrr}) are
satisfied for the so-called catenary-type function  $R(r) {=} a
\cosh{\frac{r}{a}}$, which is also even function and has the
asymptote $R(r\to \infty) \to a e^{\frac{r}{a}}$. This function
satisfies the equation $R^{''}{=}\frac{1}{a}R$ thus simplifying
the equations (\ref{eq1}) and (\ref{eq2}). We will use this radial
function below for the case of isotropic dark energy pressure with
equation of state $P{=}-\frac13 W$.

\noindent
{\it (3)}
\noindent The equation (\ref{eq1}) can be
simplified essentially, when we consider $R(r)$ to satisfy the
equation
\begin{equation}
\left(1-\frac{\kappa q_1}{R^4}\right)\frac{R''}{R}=
-\frac{\kappa(10q_1+4q_2+q_3)R'^2}{R^6} \,, \label{eq33}
\end{equation}
which follows from (\ref{eq1}) with $\sigma \equiv 1$ and
$W{+}P_{||}{=}0$. The corresponding solution can be presented in
quadratures as
\begin{equation}
Kr = \int\limits_a^R d\xi \frac{\xi^2}{\sqrt{\xi^4-a^4}}
\label{eq34}
\end{equation}
and satisfies the conditions (\ref{rrr}), when
$12q_1{+}4q_2{+}q_3{=}0$, $q_1=\frac{a^4}{\kappa}>0$, and $K$ is
arbitrary constant.

\subsection{List of models: The reference model with absent dark energy}

When the dark energy is absent, the equations (\ref{eq1}),
(\ref{eq2}) coincide with Eqs.~(25), (27) of the paper
\cite{EYM2}, if we put $R(r)= \sqrt{r^2+a^2}$, $W=0$, and
$P_{||}=0$. Let us remind that in the case of reference model we
deal with one-parameter family of exact solutions describing the
wormhole with the throat radius $a$, when the following
relationships between model parameters hold:
\begin{equation}
q_1 = \frac{a^4}{\kappa} \,, \quad
q_2=-\frac{10a^4}{3\kappa}-\frac{a^2}{6} \,, \quad
q_3=\frac{4a^4}{3\kappa}+\frac{2a^2}{3} \,. \label{zeroLambda}
\end{equation}
In fact, the product of the nonminimal parameter $q_1$ and of the
parameter $\kappa = 8\pi \nu^2$ predetermines the value of the
throat radius, thus, the wormhole does not exist when $q_1 \kappa
=0$. Since the value of the metric function $N(r)$ at $r=0$ can be
written as $N(0)= \frac13 \left(1-\frac{|\nu|}{\nu_{({\rm crit})}}
\right)$ with $\nu_{({\rm crit})} \equiv \sqrt{\frac{2
q_1}{\pi}}$, it is clear that $N(0)>0$ and the wormhole is
traversable, when  $|\nu| < \nu_{({\rm crit})}$. When the gauge
charge of the object, $\nu$, is changing (e.g., due to the charge
loss) and becomes less than the critical value $\nu_{({\rm
crit})}$, the corresponding wormhole becomes traversable. In other
words, the dimensionless parameter $\frac{\kappa}{4a^2}=
\frac{|\nu|}{\nu_{({\rm crit})}}$ is the guiding parameter of the
model. Below we will refer to these results discussing new
features of the model under consideration.

\subsection{List of models: The first model with $\omega_{||} {=} -1$}\label{modelB}

When $\omega_{||} {=} -1$, i.e., $W{+}P_{||}{=}0$, the equation
(\ref{eq1}) does not contain the metric function $N(r)$, and thus
converts into the equation for the function $\sigma$ only. Direct
integration gives us $\sigma$ as a function of $R(r)$ and its
derivative as follows:
\begin{equation}
\sigma = \sigma_0 R' \left(\frac{R^4-\kappa q_1}{R^4}
\right)^{\beta} \,, \quad \beta \equiv
\frac{10q_1{+}4q_2{+}q_3}{4q_1} \,. \label{sigma1}
\end{equation}
We are interested in the analysis of regular solutions for
$\sigma(r)$. The first regularity requirement directly follows
from (\ref{eq1}) (supplemented by the relationship
$W{+}P_{||}{=}0$): since $R^{'}(0){=}0$ but $R^{''}(0)\neq 0$, we
have to put $\left(1{-}\frac{\kappa q_1}{R^4(0)}\right){=}0$. This
regularity requirement fixes the nonminimal parameter $q_1$:
first, it has to be positive, $q_1>0$; second, it has to be
connected with the throat radius $a$ as $R(0) = a =(\kappa
q_1)^{\frac14}$. The second regularity requirement can be
explained as follows. When the throat conditions (\ref{rrr}) are
satisfied, one obtains that $R(r \to 0) \to a {+} \frac12
R^{''}(0)r^2$, and $R^{'}(r \to 0) \to R^{''}(0)r$. This means
that the function $\sigma(r)$ is regular at $r{=}0$, if and only
if $1{+}2\beta =0$, i.e., when $12q_1{+}4q_2{+}q_3=0$. As a
result, we obtain the regular solution for the metric function:
\begin{equation}
\sigma(r)= \sigma_0 \frac{R'R^2}{\sqrt{R^4-a^4}}\,, \quad
\sigma(0)=\sigma_0 \sqrt{\frac{a}{2}R^{''}(0)} \,. \label{sigma2}
\end{equation}
As for the integration constant $\sigma_0$, if we assume that
$R^{'}(r\to \infty) \to 1$, we can put $\sigma_0{=}1$, providing
$\sigma(\infty) {=} 1$.

The equation (\ref{eq2}), clearly, admits the solution with finite
value $N(0)$, when
\begin{equation}
\frac{1}{R^2(0)} = \frac{\kappa}{R^4(0)}\left[\frac12 -
\frac{q_1+q_2+q_3}{R^2(0)} \right] +  8\pi W_0\,. \label{sigma232}
\end{equation}
Thus, regularity of the metric functions $\sigma(r)$ and $N(r)$ is
possible, when three nonminimal coupling parameters $q_1$, $q_2$,
and $q_3$ are connected with the wormhole throat radius $R(0)=a$
by the following relationships:
\begin{gather}
q_1 = \frac{a^4}{\kappa} \,, \quad
q_2=-\frac{10a^4}{3\kappa}-\frac{a^2}{6}-\frac{8\pi
W_0}{3\kappa}a^{6} \,, \nonumber\\
q_3=\frac{4a^4}{3\kappa}+\frac{2a^2}{3}+\frac{32\pi
W_0}{3\kappa}a^{6} \,. \label{sigma237}
\end{gather}
When the dark energy is absent, i.e., $W_0=0$, these formulas
recover the relationships (\ref{zeroLambda}) for the nonminimal Wu-Yang wormhole
obtained in \cite{EYM2}.

Taking into account (\ref{sigma237}), we obtain immediately the
metric function $N(r)$ in the form
\begin{align}
    N(r)&=\frac{R}{R'^2(r)\sqrt{R^4-a^4}}\,\int\limits_0^r\frac{R'(x)dx}{R^2\sqrt{R^4-a^4}}\left\{R^4-a^4 \right.\nonumber\\
    {}-&\left.\frac{\kappa
    (R^2-a^2)}{2}-
    8\pi
    W_0a^6\left[\left(\frac{R}{a}\right)^{4-2\omega_\bot}-1\right]\right\}.\label{23}
\end{align}
The value $N(0)$ is finite; it is positive when
\begin{equation}
aR''(0)N(0)=\frac{1}{3}-\frac{\kappa}{12a^2}+\frac{4\pi}{3}W_0(\omega_{\bot}-2)a^{2}
>0 \,. \label{24}
\end{equation}
There are four interesting subcases.

\noindent {\it (i)} $\omega_{\bot} = -1$:

\noindent It is the subcase, when $P_{\bot}=P_{||}=-W =-W_0$, and
thus the scalars of the energy and pressure of the dark energy are
constant. We deal now with the {\it effective} cosmological
constant connected with $W_0$ by the relation $\Lambda = 8 \pi
W_0$. The metric function $N(r)$ simplifies
\begin{align}
    N(r)&=\frac{R(r)}{R'^2(r)\sqrt{R^4-a^4}}\,\int\limits_0^r\frac{R'(x)dx}{R^2(x)} \sqrt{\frac{R^2-a^2}{R^2+a^2}} \nonumber\\
    &{}
\times    \left[(R^2+a^2)(1-\Lambda R^2) -
\left(\frac{\kappa}{2}+\Lambda a^4 \right)
    \right]\,,\label{239}
\end{align}
nevertheless, the integral cannot be expressed in elementary
functions; graphs, which illustrate the behavior of the functions
of this type, will be discussed below. Since now
\begin{equation}
   aR''(0) N(0)=\frac13-\frac{\kappa}{12a^2}-\frac{\Lambda
   a^2}{2}\,, \label{249}
\end{equation}
the wormhole is traversable, when
\begin{equation}
N(0)>0  \quad \Rightarrow \quad \Lambda <
\frac{2}{3a^2}\left(1-\frac{\kappa}{4a^2}\right) \,. \label{2431}
\end{equation}

\noindent {\it (ii)} $\omega_{\bot} = 0$:

\noindent In this subcase, the dark energy manifests dust
properties in the tangential directions, $P_{\bot}{=}0$. The
corresponding metric function $N(r)$ is
\begin{align}
   N(r)&=\frac{R(r)}{R'^2(r)\sqrt{R^4-a^4}}\,\int\limits_0^r\frac{R'(x)dx}{R^2(x)} \sqrt{\frac{R^2-a^2}{R^2+a^2}} \nonumber\\
    &{}
\times    \left[(R^2+a^2)(1-8\pi W_0 a^2) - \frac{\kappa}{2}
    \right]\,.\label{2375}
\end{align}
and the traversability condition $N(0)>0$ gives
\begin{equation}
8\pi W_0 < \frac{1}{a^2}\left(1- \frac{\kappa}{4a^2}\right) \,.
\label{2475}
\end{equation}

\noindent {\it (iii)} $\omega_{\bot} = 1$:

\noindent In this case, the dark energy behaves as a stiff matter
in the tangential directions, and we obtain
\begin{align}
N(r)&=\frac{R(r)}{R'^2(r)\sqrt{R^4-a^4}}\,\int\limits_0^r\frac{R'(x)dx}{R^2(x)} \sqrt{\frac{R^2-a^2}{R^2+a^2}} \nonumber\\
    &{}
\times    \left[(R^2+a^2) - \left(\frac{\kappa}{2}+8\pi W_0
a^4\right)
    \right]\,.\label{239X}
\end{align}
thus the traversability condition yields
\begin{equation}
8\pi W_0 < \frac{2}{a^2} \left(1-\frac{\kappa}{4a^2}\right)\,.
\label{246}
\end{equation}

\noindent {\it (iv)} $\omega_{\bot} = 2$:

\noindent For this very special case, the parameter $W_0$ is
hidden,
\begin{align}
N(r)&=\frac{R}{R'^2(r)\sqrt{R^4-a^4}}\,\int\limits_0^r\frac{R'(x)dx}{R^2}\sqrt{\frac{R^2-a^2}{R^2+a^2}} \times \nonumber\\
&\times\left(R^2+a^2-\frac{\kappa}{2}\right), \label{2381}
\end{align}
and the traversability condition reads $a^2>\frac{\kappa}{4}$.

\subsection{List of models: The second model with $\omega_{||} =\omega_{\bot} =
-\frac13$}\label{modelC}

This model relates to the isotropic dark energy with the equation
of state $W+3P=0$; such model was indicated in \cite{StrGas} as a
``string gas''. In the cosmological context the condition $W+3P=0$
leads to the requirement that the acceleration parameter $q(t)$ is
equal to zero identically, and the scale factor is a linear
function of the cosmological time. Let us study this model in the
context of the wormhole  structure analysis.

When $P=-\frac13 W$, the energy density of the dark energy is
distributed according to the formula
\begin{equation}
    W(r)=W_0  \ \sigma^2 N \,,
    \label{131}
\end{equation}
and the equation (\ref{eq1}) for the metric function $\sigma(r)$
happens to be decoupled from the equation (\ref{eq2}), which
describes the metric function $N(r)$. Indeed, the equation
(\ref{eq1}) converts now to the Bernoulli differential equation:
\begin{align}
\sigma'&=\frac{\sigma}{RR'(R^4-\kappa
q_1)}\left[RR''(R^4-\kappa q_1)\right.\nonumber\\
&\left.{}+(10q_1+4q_2+q_3)\kappa
    R'^2+\frac{8\pi}{3}W_0R^6\sigma^2\right]\,,
    \label{132}
\end{align}
the formal solution to which has the form
\begin{align}
    \sigma&=R'\left(1-\frac{\kappa
    q_1}{R^4}\right)^\beta \times \nonumber\\
    &\times\left[C-\frac{16\pi W_0}{3}\int dr RR'\left(1-\frac{\kappa
    q_1}{R^4}\right)^{(2\beta-1)}\right]^{-1/2} \,,\nonumber\\
    \beta&=\frac{10q_1+4q_2+q_3}{4q_1} \,.
    \label{135}
\end{align}
In this paper, we discuss (as an illustration) only one regular
solution of this type with requirements $q_1{=}0$,
$4q_2{+}q_3{=}0$ and $\frac{8\pi}{3}W_0 a^2 {+}1 {=} 0$. The last
condition relates to the negative effective cosmological constant.
This special solution is characterized by $\sigma(r) \equiv 1$ and
the radial function $R(r)$, which describes the well-known
catenary curve:
\begin{equation}
R(r) {=} a \cosh{\frac{r}{a}} \,, \ R(0) {=} a \,, \ R'(0){=}0 \,,
\ R''(0) {=} \frac{1}{a} \,. \label{195}
\end{equation}
The  metric function $N(r)$ is now the solution to the linear
equation
\begin{align}
 N'&=-\frac{N}{R'R}\left[2R'' R+ R'^2 + 8\pi W_0 R^2 \right]\nonumber\\
    &{}- \frac{1}{2R'R^5}  \left[6 \kappa q_2 - 2R^4  + \kappa R^2 \right] \,,
\label{134}
\end{align}
it can be presented as the polynomial
\begin{align}
N(z) &= \left[\frac{3\kappa q_2}{a^4} + \frac{\kappa}{2a^2} -1
\right] + z C +  \nonumber\\
&{}+ z^2 \left[\frac{6\kappa q_2}{a^4} + \frac{\kappa}{2a^2}
\right] -z^4 \frac{\kappa q_2}{a^4} \,, \label{198}
\end{align}
where $z {=} \tanh{\frac{r}{a}}$ and $C$ is an integration
constant. Let us mention that $N(-r){=}N(r)$, when $C{=}0$. If, in
addition, we choose the nonminimal coupling constant $q_2$
according to the relationship
\begin{equation}
\frac{4\kappa q_2}{a^4} = 1- \frac{\kappa}{2a^2} \,,
\label{99}
\end{equation}
we obtain that $N(r{=}\pm \infty)=1$ and $N(0){=}\frac14
\left(\frac{\kappa}{2a^2}{-}1 \right)$. In other words, the
regular wormhole with a catenary-type throat filled by the
isotropic dark energy with the equation of state
$P{=}-\frac{1}{3}W$ is traversable, when $\kappa
> 2a^2$.

\section{Causal structure of Wu-Yang wormholes with dark energy of the
$\Lambda$-type}\label{Causal}

\subsection{Search for horizons by the method of auxiliary function}

\begin{figure}
\includegraphics[width=8cm]{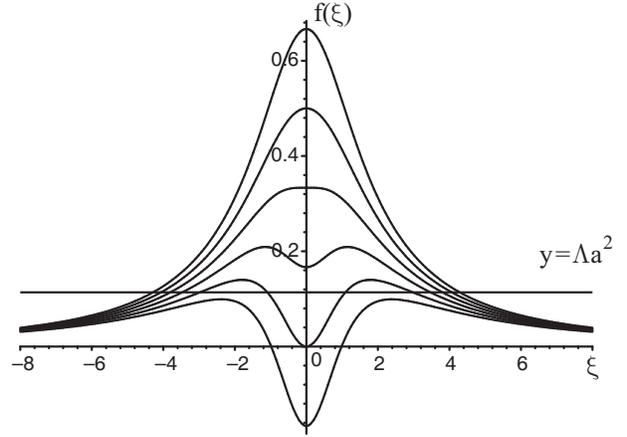}
\caption{Plots of the function
$f\left(\xi,\frac{\kappa}{a^2}\right)$ for six values of the
parameter $\frac{\kappa}{a^2}$: $0, 1, \dots, 5.$ The upper curve
relates to the minimal value of the guiding parameter,
$\frac{\kappa}{a^2}{=}0$; the maximal value of the function is
$f(0,0){=}\frac23$; this curve plays the role of separatrix. When
$\frac{\kappa}{a^2}{=}4$, $f(0,4){=}0$ according to (\ref{38}).
The lower curve corresponds to the value $\frac{\kappa}{a^2}{=}5$.
When $\frac{\kappa}{a^2}\leq 2$, the graphs have only one extremum
(maximum) at $\xi{=}0$. When $\frac{\kappa}{a^2}>2$, two symmetric
maxima and one minimum (at $\xi{=}0$) appear in the graphs. When
the parameter $\frac{\kappa}{a^2}$ increases, the maxima drift to
the left and right, respectively, and the minimal values behave as
$f\left(0,\frac{\kappa}{a^2}\to \infty \right) \to {-} \infty$.
}\label{Fig1}
\end{figure}

Let us consider the model with isotropic dark energy characterized
by the following equations of state:
\begin{equation}
P_{||}=P_{\bot}=-W=-W_0=-\frac{\Lambda}{8\pi}\,.
\end{equation}
Clearly, it is the case, which can be reduced to the nonminimal
Einstein-Yang-Mills model with cosmological constant $\Lambda$.
This submodel of the general model (of the wormhole filled with
the dark energy) is chosen to analyze in more details the problem
of horizons. To be more precise, we are interested to know how
many horizons the nonminimal Wu-Yang wormhole has and where these
horizons appear. For this purpose we fix the radial function as
\begin{equation}\label{R1}
    R(r)=\sqrt{r^2+a^2}
\end{equation}
so that the regular metric function (\ref{sigma2}) takes the form
\begin{equation}
\sigma(r) = \sqrt{\frac{r^2+a^2}{r^2+ 2a^2}} \,, \quad
\sigma(\infty) = 1 \,, \quad \sigma(0) = \frac{1}{\sqrt2} \,.
\label{sigma21}
\end{equation}
The function $\sigma(r)$ reaches neither zero nor infinite values,
thus, the causal structure of the wormhole is predetermined only
by the properties of the function $N(r)$, which can be written as
follows (see (\ref{239})):
\begin{align}
N(r)&=\frac{(r^2+a^2)^{3/2}}{r^3\sqrt{r^2+2a^2}}
\int\limits_0^r\frac{x^2
dx}{(x^2+a^2)^{3/2}\sqrt{x^2+2a^2}} \times \nonumber\\
&\times\left[{-}\Lambda x^4 {+}x^2(1{-}3\Lambda a^2)
{+}(2a^2{-}\frac{\kappa }{2}{-} 3\Lambda a^{4})\right]\,,
\label{2391}
\end{align}
\begin{equation}
   N(0)=\frac13-\frac{\kappa}{12a^2}-\frac{\Lambda
   a^2}{2}\,. \label{24955}
\end{equation}
Clearly, we deal with three-parameter family of regular solutions:
the metric function $N$ depends on the cosmological constant
$\Lambda$, on the gauge charge $\kappa \equiv 8\pi \nu^2$, and on
the parameter of the nonminimal coupling $q_1$ through the throat
radius $a \equiv (\kappa q_1)^{\frac14}$. Horizons are known to
appear at $r=r_{({\rm s})}$, where $r_{({\rm s})}$ are the zeroes
of the metric function, i.e., $N(r_{({\rm s})}){=}0$. It is
convenient to introduce the dimensionless variable
$\xi=\frac{r}{a}$ and to rewrite (\ref{2391}) as
\begin{equation}
N(\xi){=}\frac{(\xi^2{+}1)^{3/2}}{\xi^3\sqrt{\xi^2{+}2}} \
I_3(\xi)\left[f\left(\xi,\frac{\kappa}{a^2}\right){-}\Lambda a^2
\right]\,. \label{34}
\end{equation}
Here, we introduce the auxiliary function
\begin{equation}
f\left(\xi,\frac{\kappa}{a^2}\right)\equiv\frac{I_2(\xi)}{I_3(\xi)}-\frac{\kappa}{2a^2}\frac{I_1(\xi)}{I_3(\xi)}
\,, \label{37}
\end{equation}
based on the integrals
\begin{gather}
    I_1(\xi)=\int\limits_0^\xi\frac{x^2\,dx}{{(x^2+1)^{3/2}\sqrt{x^2+2}}},\\
    I_2(\xi)=\int\limits_0^\xi\frac{x^2(x^2+2)\,dx}{{(x^2+1)^{3/2}\sqrt{x^2+2}}},\\
    I_3(\xi)=\int\limits_0^\xi\frac{x^2(x^4+3x^2+3)\,dx}{{(x^2+1)^{3/2}\sqrt{x^2+2}}}
    \,.
\label{35}
\end{gather}
These integrals do not include parameters; besides, all three
functions $I_1(\xi)$, $I_2(\xi)$, $I_3(\xi)$ are odd functions of
$\xi$, and they grow monotonically for $0<\xi<\infty$. The
auxiliary function $f\left(\xi,\frac{\kappa}{a^2}\right)$ is even
with respect to the variable $\xi$, and possesses the following
asymptotical properties:
\begin{equation}
    f\left(\xi\to
    0,\frac{\kappa}{a^2}\right)=\frac23\left(1-\frac{\kappa}{4a^2}\right)\equiv
    f(0)
    \,,
\label{38}
\end{equation}
\begin{equation}
f\left(\xi\to\infty,\frac{\kappa}{a^2}\right) \sim
    \frac{3}{\xi^3}  \to 0 \,.
\label{39}
\end{equation}
The equation $N=0$, which determines the horizons, reduces now to
the equation
\begin{equation}
    f\left(\xi,\frac{\kappa}{a^2}\right)=\Lambda a^2 \,,
\label{36}
\end{equation}
thus illustrating the fact that only two effective dimensionless
parameters, $\frac{\kappa}{a^2}$ and $\Lambda a^2$, predetermine
the structure of horizons. Typical plots of the function
$f\left(\xi,\frac{\kappa}{a^2}\right)$ for several values of the
dimensionless parameter $\frac{\kappa}{a^2}$ are presented in
Fig.~\ref{Fig1}.

In order to find the number of horizons, one can determine the
number of points in which the horizontal straight line $y {=}
\Lambda a^2$ crosses the graph of the function $y {=}
f\left(\xi,\frac{\kappa}{a^2}\right)$. Clearly, depending on the
values of the parameters $\frac{\kappa}{a^2}$ and $\Lambda a^2$,
zero, one, two, three or four cross-points can appear. Let us
discuss this feature in more details.

\subsection{The structure of horizons}

In order to simplify the analysis, we distinguish three cases,
which correspond to the following values of the guiding parameter
$\frac{\kappa}{a^2}$: first, $0 \leq \frac{\kappa}{a^2}\leq 2$,
second, $2<\frac{\kappa}{a^2}\leq 4$, third, $\frac{\kappa}{a^2} >
4$.

\subsubsection{The first case: $0 \leq \frac{\kappa}{a^2}\leq 2$}

 For these values of the guiding parameter $\frac{\kappa}{a^2}$ the graph of the function
 $y=f(\xi,\frac{\kappa}{a^2})$ lies at $y > 0$ and it has one maximum at $\xi{=}0$, namely,
 $f_{{\rm max}}{=}f(0)$. The second guiding parameter, $\Lambda
 a^2$, changes, formally speaking, in the interval $-\infty <\Lambda  a^2 < \infty$.

%\vspace{3mm}
\noindent
{\it a)} When $\Lambda \leq 0$, the
horizontal straight line $y{=}\Lambda a^2$ is below the graph of
the function $y{=}f(\xi,\frac{\kappa}{a^2})$, thus, there are no
crossing. This means that $N(\xi)>0$ for arbitrary $\xi$, the
metric has no horizons and the wormhole throat is traversable.

%\vspace{3mm}
\noindent {\it b)} When $0<\Lambda a^2< f(0)$, there are two roots
of the equation $N(\xi)=0$ (symmetrical with respect to the axis
$\xi{=}0$). The metric has two horizons, which can be indicated as
the cosmological ones (see, e.g. \cite{BLZ10}), since between them
the $R$-region is situated with a traversable wormhole throat.

%\vspace{3mm}
\noindent
{\it c)} When $\Lambda a^2{=}f(0)$, we
obtain that $N(0)=0$, and two horizons coincide, being situated at
$r=0$. The wormhole throat is nontraversable.

%\vspace{3mm}
\noindent {\it d)} When $\Lambda a^2> f(0)$, the straight line
$y{=}\Lambda a^2$ lies above the graph of the function
$y{=}f(\xi,\frac{\kappa}{a^2})$. There is no crossing of these
lines, and there are no horizons. Nevertheless, $N(\xi)<0$ for
arbitrary $\xi$, i.e., now we cannot speak about a wormhole in the
standard sense of the word.

\subsubsection{The second case: $2<\frac{\kappa}{a^2}\leq 4$}

For these values of the guiding parameter $\frac{\kappa}{a^2}$,
the graph of the function  $y{=}f(\xi,\frac{\kappa}{a^2})$ lies at
$y \geq 0$, it has two symmetric maxima, $f^{*}_{{\rm max}}$, and
one minimum, at $\xi{=}0$, i.e., $f_{{\rm min}}{=}f(0)\geq 0$.

%\vspace{3mm}
\noindent {\it a)} When $\Lambda a^2 < 0$, again the straight line
$y{=}\Lambda a^2$ lies below the graph of the function
$y{=}f(\xi,\frac{\kappa}{a^2})$, $N(\xi)>0$ for arbitrary $\xi$,
the metric has no horizons, and the wormhole throat is
traversable.

%\vspace{3mm}
\noindent
{\it b)} When $0<\Lambda a^2< f_{{\rm
min}}{=}f(0)$, the equation $N(\xi){=}0$ has two roots. Again
there are two horizons of the cosmological type with intermediate
$R$-region and the traversable throat.

%\vspace{3mm}
\noindent
{\it c)} When $\Lambda a^2{=}f(0){=}0$, we
obtain $N(0){=}0$, i.e., the double horizon lies at $\xi{=}0$.

%\vspace{3mm}
\noindent
{\it d)} When $f_{\rm min}<\Lambda a^2<
f^{*}_{\rm max}$, the straight line $y{=}\Lambda a^2$ crosses the
graph of the auxiliary function in four points: $\xi{=}\pm
\xi_{\rm in}$, $\xi{=}\pm \xi_{\rm out}$. Between pairs of
horizons ($\xi_{\rm in}<|\xi|<\xi_{\rm out}$) there exist two
$R$-regions. The external horizons at $\xi{=}\pm \xi_{\rm out}$
can be indicated as the cosmological ones, while the internal
horizons at $\xi{=}\pm \xi_{\rm in}$ are the event horizons.

%\vspace{3mm}
\noindent {\it e)} When $\Lambda a^2{=} f_{\rm min}{=}f(0)$, two
internal horizons coincide; effectively, there are three horizons,
and the corresponding wormhole is nontraversable, since one
(double) horizon is situated at $\xi{=}0$.

%\vspace{3mm}
\noindent
{\it f)} When $\Lambda a^2{=} f^{*}_{\rm
max}$ the corresponding external and internal horizons coincide,
so that two horizons appear instead of four; the wormhole is
nontraversable.

%\vspace{3mm}
\noindent
{\it g)} When $\Lambda a^2> f^{*}_{\rm
max}$, again $N(\xi)$ is negative everywhere.

\subsubsection{The third case: $\frac{\kappa}{a^2} >
4$}

\begin{figure}[t]
\includegraphics[width=8cm]{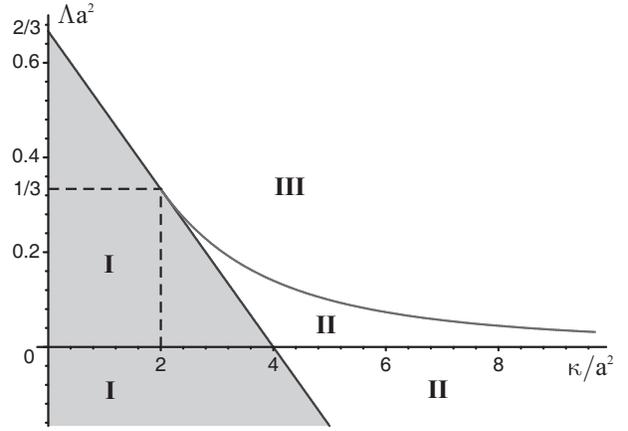} \caption{Domains on the
semiplane of the parameters $\frac{\kappa}{a^2}>0$ and  $\Lambda
a^2$, in which the $\Lambda$-influenced nonminimal Wu-Yang
wormhole has a traversable or nontraversable throats. Domain~I
(shaded) indicates wormholes with a traversable throat, i.e., when
$N(0)>0$. Domain~II relates to the spacetimes with two $R$-regions
and a $T$-region between them; the throat is nontraversable in
this case. Domain III corresponds to the spacetimes without
$R$-regions. The straight line $\Lambda
a^2{=}f(0,\frac{\kappa}{a^2}) {=}
\frac23\left(1{-}\frac{\kappa}{4a^2}\right)$ (dark) starts at
$\frac{\kappa}{a^2}{=}0$ with $\Lambda a^2 {=}\frac23$. The curved
line $\Lambda a^2{=}f^{*}_{\rm max}(\frac{\kappa}{a^2})$ (grey)
starts at $\frac{\kappa}{a^2}{=}2$ with $\Lambda a^2 {=}\frac13$;
at $\kappa{=}2a^2$, these two lines cross.}\label{Fig2}
\end{figure}

The graph of the auxiliary function has two symmetric maxima and
one minimum at $\xi{=}0$, however, now the minimal value is
negative,  $f(0)<0$. Again, the causal structure depends on the
value of the guiding parameter $\Lambda a^2$.

%\vspace{3mm}
\noindent
{\it a)} When $\Lambda a^2 <f(0)<0$, the
straight line $y{=}\Lambda a^2$ is below the graph of the
auxiliary function, and again $N(\xi)>0$ everywhere, thus, the
wormhole throat is traversable.

%\vspace{3mm}
\noindent
{\it b)} When $\Lambda a^2{=} f(0)<0$,
there is the double horizon in the throat.

%\vspace{3mm}
\noindent
{\it c)} When $f(0)<\Lambda a^2 \leq 0$,
the equation $N(\xi){=}0$ has two symmetric roots. These roots
correspond to the pair of the event horizons with a $T$-region
between them.

%\vspace{3mm}
\noindent {\it d)} When $0<\Lambda a^2< f^{*}_{\rm
max}$, there are four crossing points: $\xi{=}\pm \xi_{\rm in}$,
$\xi{=}\pm \xi_{\rm out}$. At $\xi_{\rm in}<|\xi|<\xi_{\rm out}$
there are $R$-regions, harbored by horizons. The external horizons
can be indicated as the cosmological ones, while the internal
horizons are the event ones.

%\vspace{3mm}
\noindent {\it e)} When $\Lambda a^2{=}f^{*}_{\rm max}$, the
external and internal horizons coincide.

%\vspace{3mm}
\noindent {\it f)} When $\Lambda a^2> f^{*}_{\rm
max}$, the straight line $y{=}\Lambda a^2$ does not cross the
graph of the function $y{=}f(\xi,\frac{\kappa}{a^2})$; again,
$N(\xi)<0$ for arbitrary $\xi$.

\subsection{Intermediate summary}

In order to summarize the results of the analysis we consider the
plane of the parameters $\frac{\kappa}{a^2}$, $\Lambda a^2$, and
indicate the domains without horizons, with one, two, three and
four horizons. The results are illustrated in Fig.~\ref{Fig2}.

\section{Discussion}\label{conclusion}

In this paper, we discuss new exact solutions of the wormhole type
obtained in the framework the model of nonminimal coupling between
the gauge field of the Wu-Yang type and gravity in a dark energy
environment. The solutions of this model form a seven-parameter
family: they depend generally on three nonminimal coupling
constants $q_1$, $q_2$, $q_3$ (see (\ref{sus})), on two parameters
$\omega_{||}$, $\omega_{\bot}$, which appear in the equation of
state for the dark energy (\ref{eos3}), on the charge of the gauge
field $\nu$ (see (\ref{2})), entering the master equations for the
gravity field via the parameter $\kappa \equiv 8\pi \nu^2$, and on
the initial value of the energy density of the dark energy $W_0$,
which can be transformed into the effective cosmological constant
in some special cases. In order to visualize exact solutions by
presenting the metric function $\sigma(r)$ in the explicit form
and the metric function $N(r)$ in quadratures, we considered
various models with fixed parameters $\omega_{||}$ and
$\omega_{\bot}$. Along this line we discussed the solutions with
$\omega_{||}=-1$ and various $\omega_{\bot}$ (the model with
anisotropic dark energy, see Subsection~\ref{modelB}); the
solution with $\omega_{\bot}=\omega_{||}=-\frac13$ (analog of a
string gas, see Subsection~\ref{modelC}); the solutions with
$\omega_{\bot}=\omega_{||}=-1$ (the model with effective
cosmological constant $\Lambda= 8\pi W_0$, see
Subsection~\ref{modelB}{\it (i)}).

Then, we extracted four-, three-, and two-parameter families of
{\it regular} exact solutions from the mentioned four-parameter
families by fixing the nonminimal coupling constants $q_1$, $q_2$
and $q_3$. In particular, when these constants are presented by
equations (\ref{sigma237}), the first constitutive parameter is
$\omega_{||}=-1$ and the second one, $\omega_\bot$, is arbitrary;
we find the four-parameter family of the wormhole-type solutions
with $a$ describing the throat radius, $W_0$, connected with the
initial value of the dark energy density (\ref{eos4}), $\kappa =
8\pi \nu^2$, and $\omega_\bot$ (see Subsection~\ref{modelB}). When
four guiding parameters of this anisotropic model, $a$, $W_0$,
$\kappa$, and $\omega_{\bot}$, satisfy the following inequality
$$
1+ 4\pi W_0(\omega_{\bot}-2)a^{2} > \frac{\kappa}{4a^2} \,,
$$
the throat of this wormhole is traversable, since the metric functions $\sigma(r)$ and $N(r)$ are regular, and $\sigma(0)>0$, $N(0)>0$.

Another very interesting result is obtained for the isotropic
distribution of the dark energy with
$\omega_{||}=\omega_\bot=-\frac13$ (this equation of state is
known as a ``string gas'' \cite{StrGas}). If we assume that
$q_1{=}0$, $4q_2{+}q_3{=}0$ and $\frac{8\pi}{3}W_0 a^2 {+}1 {=}
0$, the corresponding nonminimal wormhole is characterized by the
catenary-type throat profile $R(r)=a\cosh \frac{r}{a}$, by the
metric function $\sigma \equiv 1$ and by the function $N(r)$
presented explicitly by the polynomial of the fourth order of the
variable $z = \tanh{\frac{r}{a}}$ (see formula (\ref{198})). The
most illustrative regular solution of this type relates to the
one-parameter model in which $C=0$ and the additional requirement
(\ref{99}) is used in order to fix the nonminimal coupling
parameter $q_2$. This solution is of the form
$$
N(r)=\tanh^2{\frac{r}{a}} + \frac14 \left(\frac{\kappa}{2a^2}-1 \right) \cosh^{-4}{\frac{r}{a}}  \,,
$$
it has asymptotes $N(\pm \infty)=1$, and a finite value at the
center $N(0)=\frac14 \left(\frac{\kappa}{2a^2}-1 \right)$. When
$\kappa>2a^2$, the presented function $N(r)$ has no zeroes, thus,
this wormhole is traversable, and it links two regions of the
spacetime with constant negative spatial curvature (or spatially
open universes).

For the isotropic distribution of the dark energy with
$\omega_{||}=\omega_\bot=-1$, the solution (\ref{sigma21}),
(\ref{2391}) describes a symmetric magnetic wormhole joining two
asymptotically de Sitter (or asymptotically anti de Sitter)
regions with the effective cosmological constant $\Lambda=8\pi
W_0$. This solution belongs to the same class of exact solutions
describing nonminimal wormholes, which we obtained and discussed
in \cite{EYM2} (magnetic wormholes joining two asymptotically
Minkowski regions) and in \cite{BLZ10} (electric wormholes joining
an asymptotically Minkowski region and asymptotically de~Sitter
one). For this model, we considered in detail the problem of
horizons, and have shown that, depending on the values of two
effective guiding parameters $\frac{\kappa}{a^2}$ and $8\pi W_0a^2
\equiv \Lambda a^2$, the metric function $N(r)$ admits four,
three, two, one zeroes, or does not admit zeroes at all.
Correspondingly, there are two principally different wormhole
configurations in the dark energy environment, which can be
indicated as wormholes with traversable throats. First, the
wormhole can have no horizons, and thus it is traversable in the
general sense; this is possible when $\Lambda
a^2<\frac23\left(1-\frac{\kappa}{4a^2}\right)$. Second, the
wormhole can have two symmetric horizons, which are distant from
the throat. The throat of such wormhole is traversable, and two
horizons, in this sense, can be considered as the ones of a
cosmological type. All other configurations have to be indicated
as nontraversable. Indeed, when there is one zero of the metric
function $N(r)$, or there are three zeroes, one horizon is
inevitably situated at $r=0$, i.e., it appears just in the
wormhole throat. When $N(r)$ has four zeroes, all four horizons
are distant from the throat; nevertheless, this configuration has
to be indicated as the nontraversable wormhole, since, in addition
to the pair of cosmological horizons, two distant event horizons,
which close the entrance to the throat and the exit from it,
appear.

To conclude, we have to emphasize two new aspects of the obtained
results.

First, we have shown that the dark energy influence can
effectively regulate the traversability of the nonminimal Wu-Yang
wormholes: it can create the horizon just inside the throat, and
it can organize two distant horizons at the entrance and exit of
the throat. On the other hand, the dark energy can open the
nonminimal wormhole throat for traveling from one region of the
spacetime to another. For instance, when we deal with the dark
energy of the $\Lambda$-type, we can distinguish two situations:
for the case $\Lambda =0$ and $\frac{\kappa}{4a^2}>1$, the
wormhole is nontraversable; when $\Lambda \neq 0$ and $\Lambda
a^2<\frac23\left(1-\frac{\kappa}{4a^2}\right)$, the entrance to
the wormhole throat happens to be opened. When the dark energy
pressure is not constant, using the $t$-representation of the
model, we could try to find specific epochs in the Universe
history for which the cosmic gates related to fixed gauge charge
$\nu$ are opened or closed; we hope to discuss this problem in the
future.

Second, our study allows us to view from a new perspective a
situation with the number of fundamental constants. The problem
can be formulated as follows: is at least one nonminimal coupling
constant (e.g., $q_1$) a new constant of Nature, or should all
three parameters, $q_1$, $q_2$, and $q_3$, be reduced to the
combinations of already-known fundamental constants? Our study
gives food for thought in this relation. One can assume that the
nonminimal coupling constants can be connected with parameters of
the dark energy. For instance, if we assume that $\Lambda$ is
negative (it is typical for  anti-de Sitter asymptotes) and
require that the value $N(0)$ of the metric function $N(r)$ (see
(\ref{24955})) is positive and does not depend on the value of the
gauge charge $\nu$, we obtain that $q_1=-\frac{1}{6\Lambda}$ and
$N(0)=\frac13>0$. In other words, the key nonminimal coupling
constant is not independent and is reciprocal to the cosmological
constant. Clearly, this relationship is not unique, and we hope to
discuss a few interesting ideas in the next work.

\appendix

\acknowledgments This work was partially supported by the Russian
Foundation for Basic Research (Grant No. 14-02-00598), and by the
Program of Competitive Growth of Kazan Federal University (Project
No. 0615/006.15.02302.034).

\label{end}

\end{document}